\documentclass[aps, prc, amsmath, amssymb, superscriptaddress, reprint,preprintnumbers]{revtex4-2}
\usepackage[dvipdfmx]{graphicx}% to include pdf figure
\usepackage{natbib}
\usepackage{color}
\usepackage{dcolumn}% Align table columns on decimal point
\usepackage{bm}% bold math
\usepackage[T1]{fontenc}
\usepackage[utf8]{inputenc}
\usepackage{CJK}
\usepackage{url}
\usepackage[colorlinks=true,allcolors=blue,breaklinks]{hyperref}
\graphicspath{{./pic/}}

\def\ket#1{\left|{#1}\right\rangle}
\def\braket#1#2{\left\langle{#1}\middle|{#2}\right\rangle}
\def\brakket#1#2#3{\left\langle{#1}\middle|{#2}\middle|{#3}\right\rangle}

\def\nuc#1#2#3{{}^{#2}_{#3}\mathrm{#1}}
\def\urm#1{\scriptstyle{\text{\textrm{\textmd{\textup{#1}}}}}}

\def\avr#1{\left\langle{#1}\right\rangle}

\let\temp\epsilon
\let\epsilon\varepsilon
\let\varepsilon\temp
\let\temp\relax
\let\temp\phi
\let\phi\varphi
\let\varphi\temp
\let\temp\relax

\DeclareMathOperator{\sgn}{sgn}
\begin{document}
% \bibliographystyle{unsrt}
%%%%%%%%%%%%%%%%%%%%%%%%%%%%%%%%%%%%%%%%%%%%%%%%%% 
\begin{CJK*}{UTF8}{}
  \preprint{RIKEN-iTHEMS-Report-24}
  \title{Accurate and precise quantum computation of valence two-neutron systems}
  %%%%%%%%%%%%%%%%%%%%%%%%% 
  \author{Sota Yoshida (\CJKfamily{min}{吉田聡太})}
  \email{syoshida@a.utsunomiya-u.ac.jp}
  \affiliation{Institute for Promotion of Higher Academic Education, Utsunomiya University, Mine, Utsunomiya, 321-8505, Japan}
  \affiliation{School of Data Science and Management, Utsunomiya University, Mine, Utsunomiya, 321-8505, Japan}
  \affiliation{RIKEN Nishina Center for Accelerator-based Science, RIKEN, Wako 351-0198, Japan}
  %%%%%%%%%%%%%%%%%%%%%%%%% 
  \author{Takeshi Sato (\CJKfamily{min}{佐藤健})}
  \affiliation{Graduate School of Engineering, The University of Tokyo, 7-3-1 Hongo, Bunkyo-ku, Tokyo 113-8656, Japan}
  \affiliation{Photon Science Center, School of Engineering, The University of Tokyo, 7-3-1 Hongo, Bunkyo-ku, Tokyo 113-8656, Japan}
  \affiliation{Research Institute for Photon Science and Laser Technology, The University of Tokyo, 7-3-1 Hongo, Bunkyo-ku, Tokyo 113-0033, Japan}
  %%%%%%%%%%%%%%%%%%%%%%%%% 
  \author{Takumi Ogata (\CJKfamily{min}{緒方拓巳})}
  \affiliation{Graduate School of Engineering, The University of Tokyo, 7-3-1 Hongo, Bunkyo-ku, Tokyo 113-8656, Japan}
  %%%%%%%%%%%%%%%%%%%%%%%%% 
  \author{Tomoya Naito (\CJKfamily{min}{内藤智也})}
  \affiliation{RIKEN Interdisciplinary Theoretical and Mathematical Sciences Program (iTHEMS), RIKEN, 2-1 Hirosawa, Wako 351-0198, Japan}
  \affiliation{Department of Physics, Graduate School of Science, The University of Tokyo, 7-3-1 Hongo, Bunkyo-ku, Tokyo 113-0033, Japan}
  %%%%%%%%%%%%%%%%%%%%%%%%% 
  \author{Masaaki Kimura (\CJKfamily{min}{木村真明})}
  \affiliation{RIKEN Nishina Center for Accelerator-based Science, RIKEN, Wako 351-0198, Japan}
  %%%%%%%%%%%%%%%%%%%%%%%%%%%%%%%%%%%%%%% 
  \begin{abstract}
    Developing methods to solve nuclear many-body problems with quantum computers is an
    imperative pursuit within the nuclear physics community.
    Here, we introduce a quantum algorithm to accurately and precisely compute the ground state of
    valence two-neutron systems leveraging presently available Noisy Intermediate-Scale Quantum devices.
    Our focus lies on the nuclei having a doubly-magic core plus two valence neutrons in the $ p $, $ sd $, and $ pf $ shells, 
    i.e. $ \nuc{He}{6}{} $, $ \nuc{O}{18}{} $, and $ \nuc{Ca}{42}{} $, respectively.
    Our ansatz, quantum circuit, is constructed in the pair-wise form,
    taking into account the symmetries of the system in an explicit manner,
    and enables us to reduce the number of qubits and the number of CNOT gates required.
    The results on a real quantum hardware by IBM Quantum Platform show that the proposed method gives
    very accurate results of the ground-state energies,
    which are typically within $ 0.1 \, \% $ error in the energy for $ \nuc{He}{6}{} $ and $ \nuc{O}{18}{} $
    and at most $ 1 \, \% $ error for $ \nuc{Ca}{42}{} $. 
    Furthermore, our experiments using real quantum devices also show 
    the pivotal role of the circuit layout design, attuned to the connectivity of the qubits, in mitigating errors.
  \end{abstract}
  %%%%%%%%%%%%%%%%%%%%%%%%%%%%%%%%%%%%%%% 
  \maketitle
\end{CJK*}
%%%%%%%%%%%%%%%%%%%%%%%%%%%%%%%%%%%%%%%%%%%%%%%%%% 
% 
%%%%%%%%%%%%%%%%%%% 
% +++++ Intro +++++%
%%%%%%%%%%%%%%%%%%% 
% 
\section{Introduction}
\par
Electrons inside atoms, molecules, and solids, and nucleons inside atomic nuclei,
called quantum many-body systems altogether,
obey the Schr\"{o}dinger equation.
Once one was able to solve the equation,
one could, in principle, obtain most properties of such systems.
As one attempts to simulate the rich dynamics and behavior inherent in such systems,
the limitations of classical computational methods become apparent 
as the number of degrees of freedom to be considered increases.
\par
Quantum computing, with its inherent capacity to harness quantum mechanical principles for computation,
offers a promising pathway to overcome the computational bottlenecks associated with traditional methods.
While the number of qubits and the coherence time of the qubits are still limited,
which leads to the term Noisy Intermediate-Scale Quantum (NISQ) devices,
many-body systems are providing good test grounds for quantum computing, algorithms, and error mitigation techniques.
A representative algorithm to solve many-body systems is the Variational Quantum Eigensolver (VQE)~\cite{Peruzzo_14NatCom}.
In the VQE, the ground state of a given Hamiltonian is derived by minimizing the expectation value of the Hamiltonian
with respect to a parameterized trial wave function (also called the ansatz).
Afterward, many variants of the VQE algorithm have been proposed and applied to various systems,
see e.g.,~\cite{Bharti_RMP22} and references therein.
\par
The nucleus, a system composed of strongly interacting nucleons,
is a striking example of a complex many-body system that is difficult to compute.
When one tries to apply the full configuration interaction (CI) method
or valence CI method, the so-called shell-model calculation,
it is not uncommon to encounter dimensions exceeding $ 10^{15} $ (see e.g.~\cite{BARRETT_13Rev,LAUNEY_16Rev,Otsuka_22NatComm}).
To tackle such large dimensional problems, variants for CI methods,
which enable us to go beyond the current limitation of the dimension of exact diagonalization $ \approx 10^{11} $,
are proposed (e.g., 
Monte Carlo shell-model~\cite{MCSM_Rev1,MCSM_Rev2},
importance truncation scheme~\cite{Roth_09PRC}, 
symmetry-adopted methods~\cite{LAUNEY_16Rev}, 
quasiparticle vacua shell model~\cite{Shimizu_PRC21_QVSM}).
In parallel with this direction, emulators or surrogate models for CI methods,
utilizing the eigengvector continuation~\cite{Frame_PRL18,Duguet_23EC},
are proposed \cite{Konig_PLB20,Wesolowski_PRC21,Yoshida_PTEP22}.
However, it is still demanding to develop computationally more efficient methods 
to tackle cutting-edge calculations for e.g. nuclei around driplines and/or larger systems.
Thus, nuclear physics also offers a good test ground for quantum computing and algorithms.
\par
There are several pioneering applications of quantum algorithms to nuclear many-body problems,
mostly on the Lipkin model~\cite{Cervia_PRC21,Romero_PRC22,Hlatshwayo_PRC22}
and on shell-model Hamiltonians~\cite{Stetch_PRC22,Kiss_PRC22,Spain_SciRepo2023,Sarma_PRC23,Bhoy_2402.15577}.
However, this is still developing, and further research is needed to usher in the era of quantum computers for a more general class of nuclear many-body problems.
\par
This study explores how to solve the valence two-neutron systems accurately and precisely with the present NISQ devices.
By taking into account the symmetries of the system in an explicit manner,
and selecting appropriate ansatz (circuits), the optimization method,
and the error mitigation technique,
we will show that the proposed method is proven to give very accurate results
of the ground-state energies with fewer CNOT gates than the previous works
utilizing e.g.,~Unitary Coupled Cluster (UCC) ansatz and its variants~\cite{Stetch_PRC22,Kiss_PRC22,Spain_SciRepo2023,Sarma_PRC23,Bhoy_2402.15577}.
\par
The target systems are the nuclei having doubly-magic core plus two valence neutrons,
such as $ \nuc{He}{6}{} $, $ \nuc{O}{18}{} $, and $ \nuc{Ca}{42}{} $.
Within the so-called $M$-scheme, even nuclei are described by only the configurations coupled to $ M \equiv \sum j_z = 0 $,
where $ j_z $ is the projection of the total angular momentum $ j $ on the $ z $-axis.
Besides, the $ J = 0 $ states of two neutrons are expressed as the superposition of the configurations with a time-reversal pair of nucleons.
Accordingly, the creation (annihilation) operators of two nucleons are combined into
one creation (annihilation) operator in the second quantized form,
which we call the ``pair-wise'' form.
For the two-protons systems or two-hole systems, the same procedure can be applied.
With the help of the mapping of nucleon operators to the pair-wise operators,
one can significantly reduce the number of CNOT gates needed to prepare wave functions and measure energies.
\par
Using IBM Quantum (IBMQ) devices, we will show that the proposed method gives
accurate and precise results of the ground-state energies
of the target systems with % 1 \, \%$ error at most.
It is also shown that the circuit layout design, attuned to the connectivity of the qubits,
plays a pivotal role in mitigating errors.
\par
This paper is organized as follows.
Section \ref{sec:Method} is devoted to explain the basics of shell-model calculations and the adopted circuit to encode the problems onto
quantum circuits.
The results on both simulators and real devices for the target two-neutron systems are shown in Sec.~\ref{sec:Results}.
Then, the summary follows in Sec.~\ref{sec:Summary}.
Implementation and experiments for this work are performed using the Qiskit~\cite{Qiskit} and IBM Quantum Platform~\cite{IBM_Quantum}.
%
%%%%%%%%%%%%%%%%%%%%%%%%% 
% +++++ METHODOLOGY +++++%
%%%%%%%%%%%%%%%%%%%%%%%%% 
% 
\section{Methodology}
\label{sec:Method}
\subsection{Shell-model calculation}
\par
The shell model Hamiltonian is written as
\begin{equation}
  \label{eq:Hsm}
  H
  =
  \sum_i
  \epsilon_i
  \hat{a}^{\dag}_i \hat{a}_i
  +
  \frac{1}{4}
  \sum_{ijkl}
  V_{ijkl}
  \hat{a}^{\dag}_i \hat{a}^{\dag}_j \hat{a}_l \hat{a}_k,
\end{equation}
where $ i $ denotes the single-particle state having $ \left\{ n, l, j, j_z, t_z \right\} $,
$ \epsilon_i $ is the single-particle energy,
and $ V_{ijkl} $ is the two-body matrix element.
Here, $n$ is the principal quantum number, $l$ is the orbital angular momentum,
$j$ is the total angular momentum,
$j_z$ is the projection of $j$ on the $z$-axis,
and $t_z$ is the isospin projection.
When we refer to merely orbital, it means the single-particle state classified by $ \left\{ n, l, j, j_z, t_z \right\} $,
and the term ``jj-coupled orbital'' will be used to refer to the single-particle states having the same $ \left\{ n, l, j, t_z \right\} $.
If one let the summation to be \textit{canonically-ordered}
\footnote{
  We call the summation to be canonically-ordered
  if the summation is done in the following order:
  $ i \leq j $, $ k \leq l $, $ i \leq k $, and $ j \leq l $ if $ i = j $.
}
and the factor
$ \sqrt{\left( 1 + \delta_{ij} \right) \left( 1 + \delta_{kl} \right)} $
for proton-proton and neutron-neutron interactions
is absorbed into the definition of $ V_{ijkl} $, the factor of $ 1/4 $ can be omitted in numerical calculations.
\par
In shell-model calculations, the Hamiltonian is given in the adopted model space
and diagonalized in the model space by means of e.g., Lanczos method.
In this work, we use such exact diagonalization results as the reference values to
be compared with the results of the quantum calculation.
\par
The systems of interest are the ground states of two-neutron systems within the $ p $, $ sd $, and $ pf $ shell,
i.e., $ \nuc{He}{6}{} $, $ \nuc{O}{18}{} $, and $ \nuc{Ca}{42}{} $, respectively.
Table~\ref{tab:exact} summarizes the exact results for these nuclei with phenomenological effective interactions,
Cohen-Kurath ($ p $ shell)~\cite{
  CohenKurath},
USDB ($ sd $ shell)~\cite{
  USDB},
and GXPF1A ($ pf $ shell)~\cite{
  GXPF1A}.
One can obtain the corresponding interaction files and reproduce these results with shell-model codes such as \textsc{KSHELL}~\cite{
  KSHELL1,
  *KSHELL2}
and NuclearToolkit.jl~\cite{
  NuclearToolkit.jl,
  *Repo_NuclearToolkit.jl}.
For these nuclei, the calculation with a classical computer is easy, but it is still a good test ground for the quantum computer and error mitigation techniques.
\begin{table}[b]
  \begin{ruledtabular}
    \caption{Exact ground-state energies for the target nuclei and interactions.}
    \label{tab:exact}
    \begin{tabular}{lld}
      Nucleus & Interaction & \multicolumn{1}{c}{$ E_{\urm{g.s.}} $ ($ \mathrm{MeV} $)} \\
      \hline
      $ \nuc{He}{6}{} $  & Cohen-Kurath (ckpot)~\cite{CohenKurath} &  -3.90981 \\
      $ \nuc{O}{18}{} $  & USDB~\cite{USDB}         & -11.93179 \\
      $ \nuc{Ca}{42}{} $ & GXPF1A~\cite{GXPF1A}     & -19.73368 
    \end{tabular}
  \end{ruledtabular}
\end{table}
\subsection{Pair-wise form of shell model Hamiltonian}
\par
For two-neutron systems, the states having $ J = 0 $ are expressed as
the superposition of the configurations with a time-reversal pair of nucleons.
By mapping those \textit{pair-wise} configurations to the qubit states,
the wave function and Hamiltonian can be written with less number of
qubits than the number of single-particle states.
\par
Let us introduce the following pair creation, pair annihilation, and pair occupation number operators:
\begin{subequations}
  \begin{align}
    A^{\dag}_{\tilde{i}}
    & =
      c^{\dag}_i c^{\dag}_{\bar{i}}, \\
    A_{\tilde{i}}
    & =
      c_{\bar{i}} c_i,\\
    N_{\tilde{i}}
    & =
      c^{\dag}_i c_i
      +
      c^{\dag}_{\bar{i}} c_{\bar{i}}. \label{eq:opN}
  \end{align}
\end{subequations}
Here, $ \bar{i} $ denotes the time-reversed state of $ i $,
and let a single index $ \tilde{i} $ denote the pair $ i $ and $ \bar{i} $ ($ > i $).
Note that we assume that the index $ i $ in the Hamiltonian~\eqref{eq:Hsm} is assigned
to the single-particle state with harmonic oscillator quanta,
$ \left\{ n, l, j, j_z, t_z \right\} $, 
and is indexed in ascending order of $ n $, $ l $, $ j $ and $ j_z $.
These operators satisfy the following relations~\cite{
  Khamoshi2021J.Chem.Phys.154_074113}:
\begin{subequations}
  \begin{align}
    \left[ A_{\tilde{i}}, A^{\dag}_{\tilde{j}} \right]
    & =
      \delta_{\tilde{i} \tilde{j}} \left( 1 - N_{\tilde{i}} \right), \\
    \left[ N_{\tilde{i}}, A^{\dag}_{\tilde{j}} \right]
    & =
      2
      \delta_{\tilde{i}\tilde{j}} A^{\dag}_{\tilde{j}}.
  \end{align}
\end{subequations}
\par
If one considers only the pair-wise (pw) configurations, 
the Hamiltonian Eq.~\eqref{eq:Hsm} can be written as
\begin{equation}
  \label{eq:Hpw}
  H^{\urm{pw}}
  =
  \sum_i
  \bar{h}_i
  A^{\dag}_i A_i
  +
  \sum_{i \leq j}
  \bar{V}_{ij}
  A^{\dag}_i A_j,
\end{equation}
where the tilde on the index is omitted for simplicity.
Each term has the following relations with the original Hamiltonian:
\begin{subequations}
  \begin{align}
    \sum_i
    \bar{h}_i
    A^{\dag}_i A_i
    & =
      \sum_i
      \left(
      \epsilon_i
      a^{\dag}_i a_i
      +
      \epsilon_{\bar{i}}
      a^{\dag}_{\bar{i}} a_{\bar{i}}
      \right), \\
    \sum_{i \leq j}
    \bar{V}_{ij}
    A^{\dag}_i A_j
    & =
      \sum_{i \leq j}
      V_{i \bar{i} j \bar{j}}
      a^{\dag}_i a^{\dag}_{\bar{i}} a_{\bar{j}} a_{j},
  \end{align}
\end{subequations}
where $ i < \bar{i} $ is assumed in the right-hand sides of the above equations.
To encode the Hamiltonian of the target system, Eq.~\eqref{eq:Hpw}, into the quantum circuit,
one needs to transform the Hamiltonian into the form of Pauli operators.
The transformed form of Eq.~\eqref{eq:Hpw} is to be~\cite{Khamoshi2020QuantumSci.Technol.6_014004}
\begin{equation}
  \label{eq:Hpw_qubit}
  H^{\urm{pw}}_{\urm{qubit}}
  =
  \sum_i
  \frac{\bar{h}_i + \bar{V}_{ii}}{2} \left( I_i - Z_i \right)
  +
  \sum_{i < j}
  \bar{V}_{ij} \left( X_i X_j + Y_i Y_j \right),  
\end{equation}
where $I_i$, $X_i$, $Y_i$, and $Z_i$ are, respectively, the identity operator, Pauli $X$, $Y$, and $Z$ operators acting on $i$-th qubit. 
\subsection{Circuit for wave functions}
\par
Here in this subsection, we introduce our ansatz, the circuit for the wave function of valence two-neutron systems.
In the following, we restrict ourselves to the cases of $ \nuc{He}{6}{} $ and $ \nuc{O}{18}{} $,
but it is straightforward to extend to $ \nuc{Ca}{42}{} $ or other nuclei having either two neutrons or two protons in the valence space.
\par
Throughout this work, we work on two-neutron systems and 
the qubits correspond to not the single-particle states but the pair-wise configurations.
For the $ \nuc{He}{6}{} $, the valence space is $ 0p_{1/2} $ and $ 0p_{3/2} $,
and the total number of the pair-wise configurations is three.
The qubits are assigned to the pair-wise configurations 
in the ascending order of $ l $ and $ j $, and the descending order of $ \left| j_z \right| $.
For the $ \nuc{He}{6}{} $ case,
the three qubits, $ \ket{0}_0 $, $ \ket{0}_1 $, and $ \ket{0}_2 $, are assigned to the pair-wise configurations of 
$ 0p_{1/2} \left( \left| j_z \right| = 1/2 \right) $,
$ 0p_{3/2} \left( \left| j_z \right| = 3/2 \right) $,
and
$ 0p_{3/2} \left( \left| j_z \right| = 1/2 \right) $,
respectively.
Similarly, the six qubits for $ \nuc{O}{18}{} $ on the $ sd $ shell are assigned to the neutron pairs in 
$ 1s_{1/2} $,
$ 0d_{3/2} \left( \left| j_z \right| = 3/2, \, 1/2 \right) $,
and 
$ 0d_{5/2} \left( \left| j_z \right| = 5/2, \, 3/2, \, 1/2 \right) $, respectively.
In a similar manner, the ten qubits are needed for the $\nuc{Ca}{42}{}$ case.
\begin{figure}[tb]
  \centering
  \includegraphics[width=0.6\linewidth]{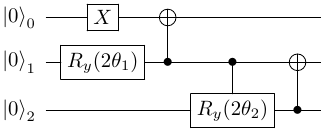}
  \caption{
    Circuit for $ \nuc{He}{6}{} $.
    The three qubits,
    $ \ket{0}_0 $, $ \ket{0}_1 $, and $ \ket{0}_2 $,
    are assigned to the pair-wise configurations of
    $ 0p_{1/2} \left( \left| j_z \right| = 1/2 \right) $,
    $ 0p_{3/2} \left( \left| j_z \right| = 3/2 \right) $,
    and
    $ 0p_{3/2} \left( \left| j_z \right| = 1/2 \right) $, respectively.}
  \label{fig:circuit_He6}
\end{figure}
\begin{figure}[tb]
  \centering
  \includegraphics[width=1.0\linewidth]{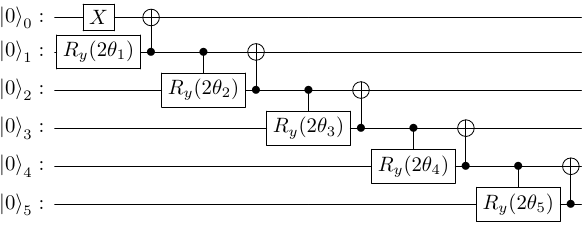}
  \caption{
    Circuit for $ \nuc{O}{18}{} $.
    The six qubits,
    $ \ket{0}_0 $, $ \ket{0}_1 $, \ldots, $ \ket{0}_5 $,
    are assigned to the pair-wise configurations of
    $ 1s_{1/2} $,
    $ 0d_{3/2} \left( \left| j_z \right| = 3/2, \, 1/2 \right) $,
    $ 0d_{5/2} \left( \left| j_z \right| = 5/2, \, 3/2, \, 1/2 \right) $, respectively.}
  \label{fig:circuit_O18}
\end{figure}
\par
The circuits for the $ \nuc{He}{6}{} $ and $ \nuc{O}{18}{} $ are shown, respectively,
in Figs.~\ref{fig:circuit_He6} and \ref{fig:circuit_O18}.
One can realize the particle number conserving wave function with
the combination of CNOT and $ R_y $ gates like these circuits.
While the qubits in a circuit are ordered from the right to the left in Qiskit~\cite{Qiskit},
 we express the qubits in the same order as the corresponding circuit,
 i.e., the bitstring ``100'' corresponds to the state $ \ket{1}_0\otimes \ket{0}_1 \otimes \ket{0}_2 $.
With the number of pair-wise configurations $ N $ and the fact that 
the controlled-$ R_y $ gate is decomposed into two CNOT gates and $ R_y $ gates,
the number of CNOT gates required for the ansatz is $ 3N - 5 $.
This value is much smaller than the number of CNOT required for e.g., the UCC ansatz.
In terms of the number of CNOTs alone, it is possible to further reduce the number of CNOTs
by using the technique proposed by Ref.~\cite{Yordanov2020pra},
but for the sake of simplicity, a simple circuit will be used in this work.
\subsection{Measurement of energy}

\par
The expectation value of the first term of Eq.~\eqref{eq:Hpw_qubit},
which contains only the $ I $ and $ Z $ operators,
can be obtained by the measurements of the ansatz circuit.
On the other hand, another circuit is required for evaluating the second term.
We consider two methods (A and B) of the measurement whose difference mainly lies
in the measurement of the second term of Eq.~\eqref{eq:Hpw_qubit}.
\par
The method A is to directly evaluate the expectation value of
the Pauli operators such as $I, Z$ and ones associated with $ XX $ and $ YY $ terms.
To this end, one can measure the ansatz circuit plus the gates operating $ XX $ or $ YY $ over two qubits.
\par
On the other hand, the method B is based on the so-called computational basis sampling technique,
which was introduced in Ref.~\cite{Kohda_Qunasys_PRR22}.
Under this method, the expectation value of the Hamiltonian is given as
\begin{align}
  \brakket{\psi}{H}{\psi}
  & =
    \sum_{m,n=0}^N
    \braket{\psi}{m}
    \brakket{m}{H}{n}
    \braket{n}{\psi}
    \notag, \\
  & = 
    \sum_{m,n=0}^N
    \left| \braket{m}{\psi} \right|^2
    \left| \braket{n}{\psi} \right|^2
    \frac{\brakket{m}{H}{n}}{\braket{m}{\psi} \braket{\psi}{n}},
    \label{eq:Hexpec}
\end{align}
where $ N $ is the number of configurations. 
The first two factors in the above equation,
$ \left| \braket{m}{\psi} \right|^2 $ and $ \left| \braket{n}{\psi} \right|^2 $,
correspond to the amplitudes of each configuration $ m $ and $ n $, e.g., $ \ket{100} $ for $ \nuc{He}{6}{} $.
As a post-processing step, one can remove the configurations violating particle number conservation, such as $ \ket{110} $ for $ \nuc{He}{6}{} $. Hence this gives variational estimates.
These amplitudes and the denominator of the third factor,
$ \braket{m}{\psi} \braket{\psi}{n} $,
are evaluated by measurements of the quantum circuits,
and the numerator $ \brakket{m}{H}{n} $ is evaluated in a classical manner.
More precisely, we evaluate the expectation value for $ XX $ and $ YY $ terms through
the sign of those terms:
\begin{align}
  \brakket{\psi}{X_j \otimes X_k}{\psi}
  & =
    \sqrt{\sigma^2_j \sigma^2_k} \sgn \left[ \sigma_j \sigma_k \right], \\
  & =
    \sqrt{\avr{N_j} \avr{N_k}} \sgn \left[ \avr{X_j X_k} \right]
    \label{eq:XXsign}
\end{align}
where $ \sigma_i \equiv \braket{0 \ldots 1_i \ldots 0}{\psi} $ is the projector onto the $ i $-th qubit,
and $ N_i = |\sigma_i|^2 $ is the pair-occupation number operator defined in Eq.~\eqref{eq:opN}.
Besides the ansatz circuit to create trial wave functions, one needs to prepare another circuit to measure the sign of $ XX $ term.
This can be achieved by measuring the circuit with Hadamard gates added to all qubits of the ansatz circuit.
While the method A gives energy estimates as the mean value of measurements, the method B, which is inherently variational,
gives the energy estimate as the minimum one among the measurements.
%
%%%%%%%%%%%%%%%%%%%%% 
%%%% +++++ Results +++++ %%%%
%%%%%%%%%%%%%%%%%%%%% 
%
\section{Results\label{sec:Results}}
\par
Here in this section, we show the results of the quantum computation of the ground-state energies of $ \nuc{He}{6}{} $, $ \nuc{O}{18}{} $, and $ \nuc{Ca}{42}{} $ using the proposed ansatz and methods.
\par
Since the nuclei of interest are two-neutron systems in the valence space,
one can determine the angle of the $ R_y $ gate by diagonalizing the Hamiltonian of Eq.~\eqref{eq:Hpw} directly by classical computers.
Throughout this study,
we restrict ourselves to the case that the circuit parameters are fixed 
to such ones giving the exact ground-state energies.
The exception can be found in Appendix.~\ref{sec:Optim} discussing the optimization of the circuit parameters starting from random initial values.
\par
Experiments are performed using the noise-free simulators (denoted as sim.~FTQC),
noisy simulators mimicking IBMQ devices (sim.~NISQ), and real quantum device from IBM Quantum Platform (Real).
We use the \textit{ibm\_brisbane} device having 127 qubits for sim.~NISQ/Real calculations.
For the results with the simulators, the number of shots is set to $ 100,000 $,
and the number of shots for the real device is set to $ 20,000 $ throughout this study.

\begin{figure}%[tb]
  \centering
  \includegraphics[width=1.0\linewidth]{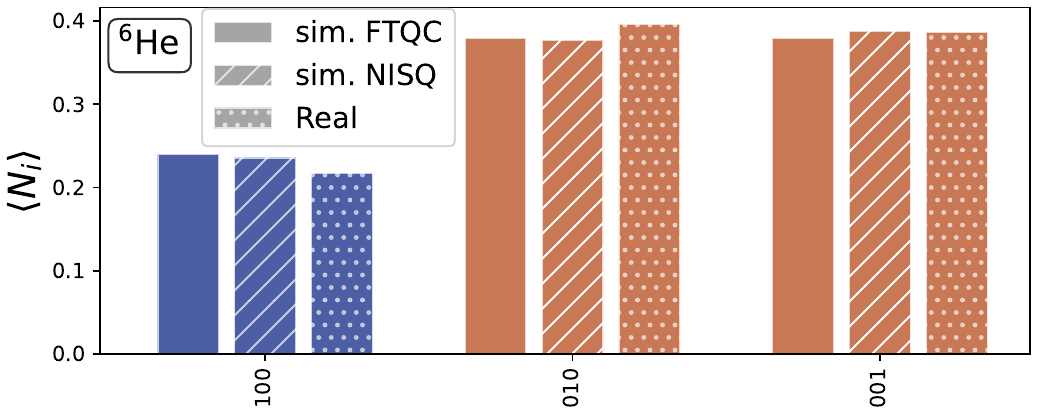}
  \caption{
    The amplitude of each configuration 
    $ \langle N_i \rangle \equiv |\langle 0..1_i..0|\psi \rangle|^2$ for $ \nuc{He}{6}{} $.
    Each color corresponds to the pair-wise configurations having the same $ \left\{ n, l, j, t_z \right\} $ quanta,
    $ 0p_{1/2} $ (blue) and $ 0p_{3/2} $ (orange).
    Note that the expectation values for the sim.~NISQ/Real cases are re-normalized after the removal of the configurations violating the particle number conservation.}
  \label{fig:He6_psi2}
  % \end{figure}
  % \begin{figure}
  \centering
  \includegraphics[width=1.0\linewidth]{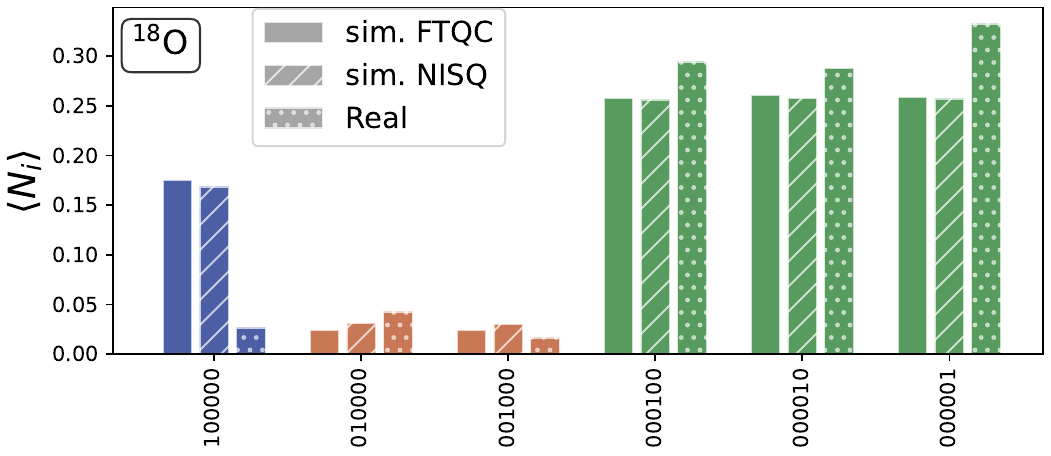}
  \caption{
    The counterpart of Fig.~\ref{fig:He6_psi2} for $ \nuc{O}{18}{} $.
    The colors correspond to
    $ 1s_{1/2} $ (blue),
    $ 0d_{3/2} $ (orange),
    and $ 0d_{5/2} $ (green) orbitals.}
  \label{fig:O18_psi2}
  % \end{figure}
  % \begin{figure}
  \includegraphics[width=1.0\linewidth]{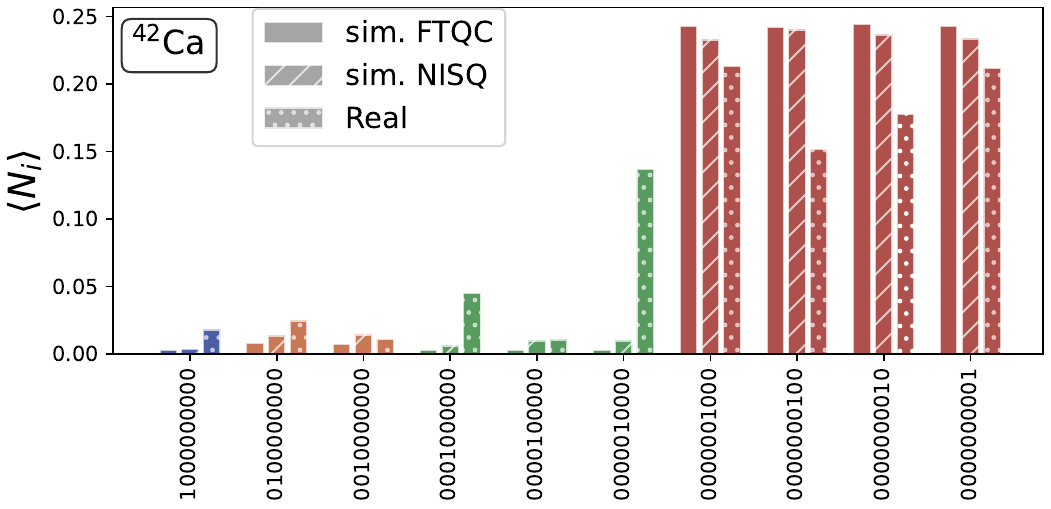}
  \caption{
    The counterpart of Fig.~\ref{fig:He6_psi2} for $ \nuc{Ca}{42}{} $.
    The different colors correspond to
    $ 1p_{1/2} $ (blue),
    $ 1p_{3/2} $ (orange),
    $ 0f_{5/2} $ (green),
    and $ 0d_{7/2} $ (red) orbitals.
  }
  \label{fig:Ca42_psi2}
\end{figure}

\subsection{Measurements of pair occupation numbers}

\par
In what follows, we show the amplitudes of each configuration,
i.e. the expectation value of pair occupation number Eq.~\eqref{eq:opN},
obtained as a result of the measurements of the ansatz circuits.
In Figs.~\ref{fig:He6_psi2}--\ref{fig:Ca42_psi2},
the relative amplitude of each configuration is shown for $ \nuc{He}{6}{} $,
$ \nuc{O}{18}{} $, and $ \nuc{Ca}{42}{} $, respectively.

\par
From noise-free results (sim.~FTQC) in these figures,
one can see that the expectation values of the states occupying 
the same jj-coupled orbital are identical within a statistical error.
This sim.~FTQC results is by construction identical to the exact results.
As a whole, ground-state wave functions of the nuclei of interest concentrate on
the configurations occupying the lowest jj-coupled orbital,
which is usually to be the orbital with the highest angular momentum $ j $ among
the adopted valence orbitals, i.e., $ 0d_{5/2} $ ($ 0f_{7/2} $) for the $ sd $ ($ pf $) shell.

On the other hand, the state preparation on the noisy simulators (sim.~NISQ) and the real device (Real)
show certain deviations from the exact results.
It should be noted that sim.~NISQ and Real results may vary depending on the quantum device employed and its calibration status.

Here we discuss how the noise affects the results from the viewpoint of particle number conservation.
Since we now consider two neutron systems and their pair-wise configurations,
the ideal result is to observe the configurations in which only a single bit is set to one.
The noise-free simulator, of course, gives the expected results,
and even with the noisy simulator, we get particle number-conserved configurations for roughly $ 80 $--$ 90 \, \% $ of all shots.
However, the probability to obtain particle number conserving configuration on \textit{ibm\_brisbane}
drops to $ \approx 60 \, \% $ for measurements of $ \nuc{O}{18}{} $ 
and it becomes even worse, $ \approx 30 \, \% $, for $ \nuc{Ca}{42}{} $.

In the Table~\ref{tab:PercentNocc}, we summarize the observed distribution of the particle numbers in the results on the \textit{ibm\_brisbane} device.
The $ \bar{N}_{\urm{occ}} $ denotes the average number of the pair occupation number among five different experiments.
As a general trend, the configurations with $ \bar{N}_{\urm{occ}} = 2 $ and $ N_{\urm{occ}} = 3 $, in that order, are observed.
Various factors such as decoherence, cross talk, gate errors and readout errors are expected to contribute to the deviations.
In Sec.~\ref{sec:Error}, we will introduce the error mitigation technique modifying the circuits
to improve the ratio to obtain the configurations with $ N_{\urm{occ}} = 1 $.

\begin{table*}[tb]
  \begin{ruledtabular}
    \caption{The breakdown of the configurations measured on a real quantum device.
      The $ \bar{N}_{\urm{occ}} $ denotes the average of the pair occupation numbers obtained through five independent experiments on the \textit{ibm\_brisbane} device.
    The values in parentheses are the results with slightly modified circuits.
    See Sec.~\ref{sec:Error} for more details.}
    \label{tab:PercentNocc}
    \begin{tabular}{lccccc}
      & \multicolumn{5}{c}{Percentile ($ \% $)} \\
      \cline{2-6}
      Nucleus & $ \bar{N}_{\urm{occ}} = 1 $ (target) 
              & $ \bar{N}_{\urm{occ}} = 0 $ & $ \bar{N}_{\urm{occ}} = 2 $ & $ \bar{N}_{\urm{occ}} = 3 $ & others \\
      \hline
      $ \nuc{He}{6}{} $ & $ 92 $          & $ 2 $         & $  6 $          & $ < 1 $         & N/A \\     
      $ \nuc{O}{18}{} $ & $ 63 $ ($ 90 $) & $ 3 $ ($ 3 $) & $ 29 $ ($  4 $) & $  4 $ ($  3 $) & $ < 1 $ \\
      $ \nuc{Ca}{42}{}$ & $ 30 $ ($ 53 $) & $ 8 $ ($ 3 $) & $ 38 $ ($ 28 $) & $ 18 $ ($ 11 $) & $ < 5 $ \\
    \end{tabular}
  \end{ruledtabular}   
\end{table*}

\subsection{Measurements of ground-state energies: noise-free simulator (sim.~FTQC) case}
\begin{figure*}[tb]
  \centering
  \includegraphics[width=1.0\linewidth]{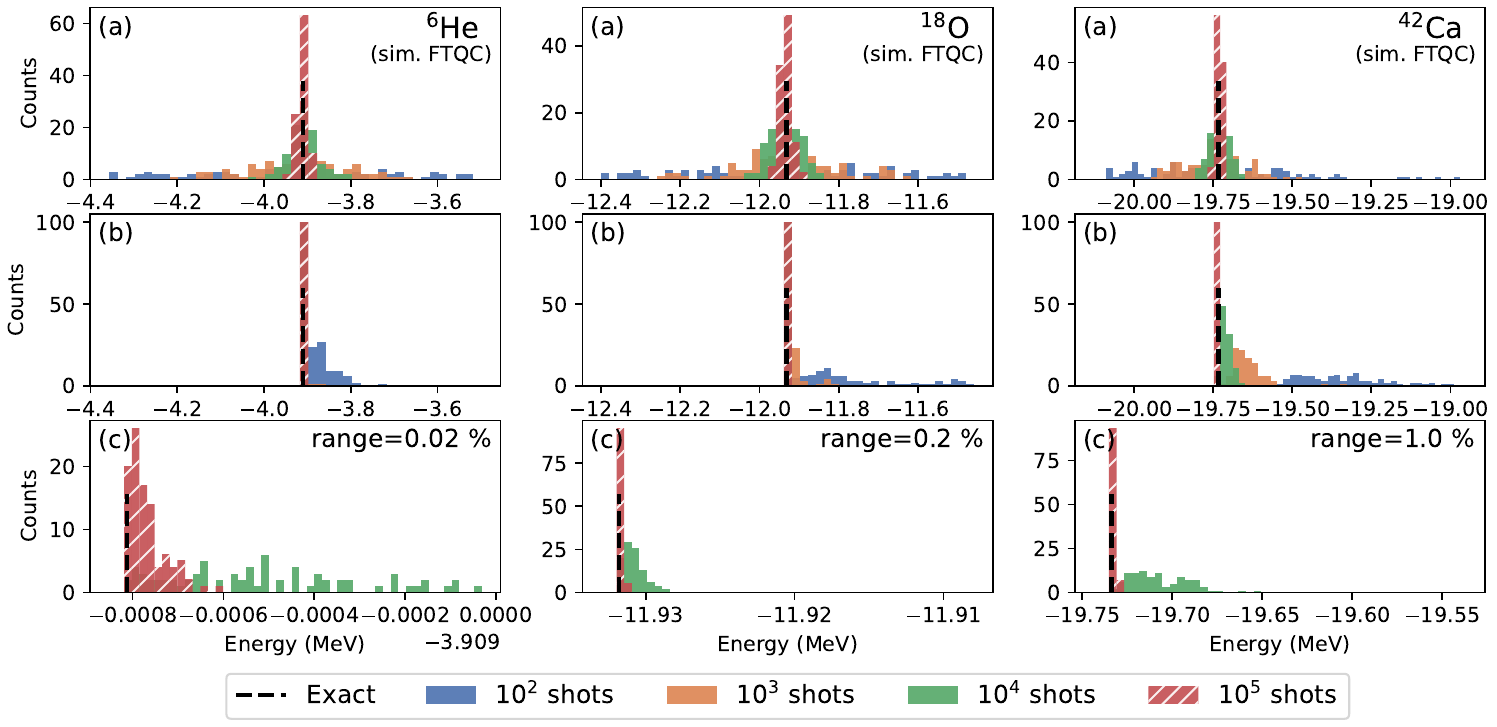}
  \caption{
    Energy estimation of ground states of $ \nuc{He}{6}{} $, $ \nuc{O}{18}{} $, and $ \nuc{Ca}{42}{} $ by noise-free simulators.
    (a)~histogram of the energy estimation by method A,
    (b)~histogram of the energy estimation by method B,
    and (c)~enlarged view of panel (b) around the exact ground-state energy.
    The percentile in panel (c) means the window of the energy range with respect to the $ E_{\urm{exact}} $.
    In panels (a) and (b), the bin width is set to 20 keV, while 50 bins are used in panel (c) within the energy range given by the percentiles.}
  \label{fig:hist_FTQC}
  \centering
  \includegraphics[width=1.0\linewidth]{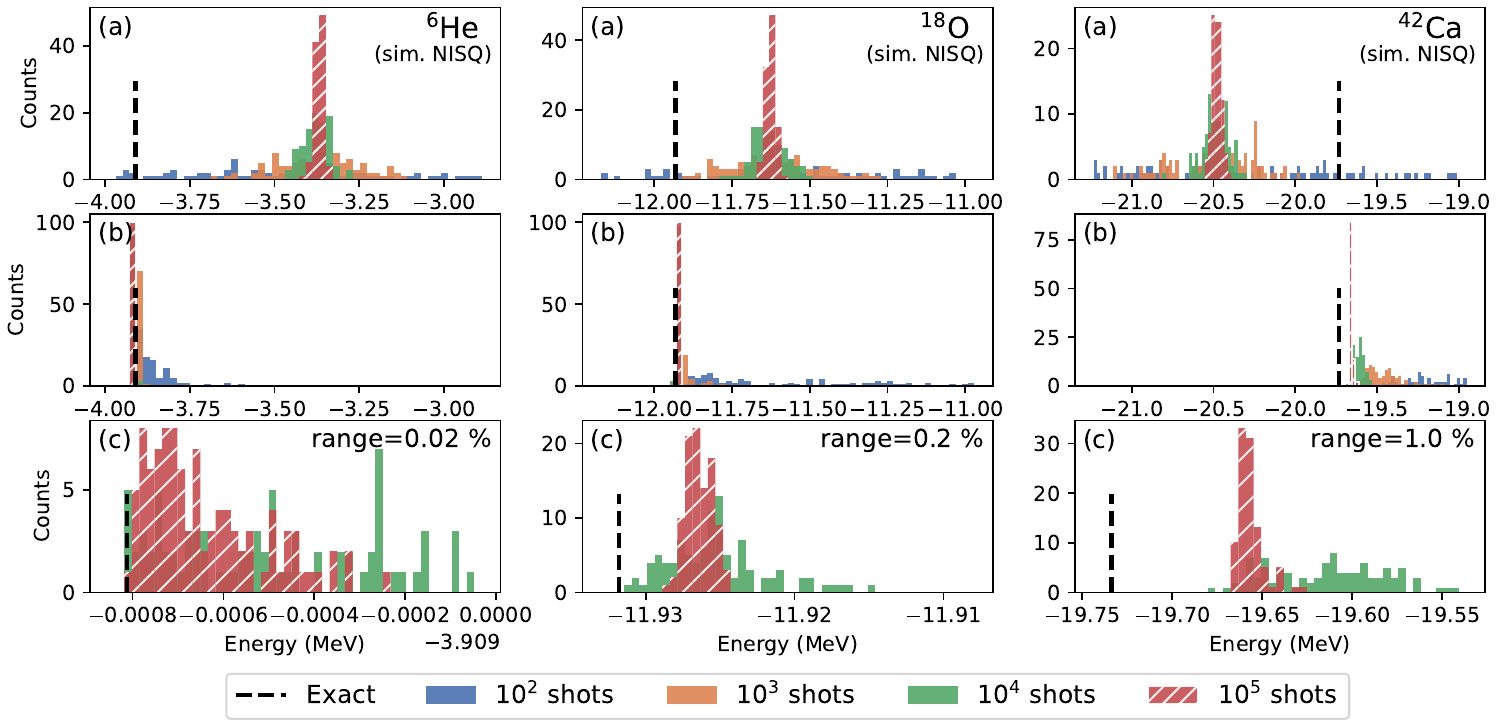}
  \caption{
    Energy estimation of ground states of $ \nuc{He}{6}{} $, $ \nuc{O}{18}{} $, and $ \nuc{Ca}{42}{} $ by noisy simulators
    mimicking the \textit{ibm\_brisbane} device.
    The bin width and meaning of the panels and percentiles are the same as in Fig.~\ref{fig:hist_FTQC}.
    }
  \label{fig:hist_NISQ}
\end{figure*}
\par
In Fig.~\ref{fig:hist_FTQC}, the sim.~FTQC results of measured ground-state energy are shown by the histograms for 100 different runs.
From left to right, the target nucleus is $ \nuc{He}{6}{} $, $ \nuc{O}{18}{} $, and $ \nuc{Ca}{42}{} $.
The results of method A and method B are shown in the panels (a) and (b), respectively.
The panels (c) show the enlarged views of the panels (b) around the exact result.
The percentile in the panels (c) means the window of the energy range with respect to the exact ground-state energy.
\par
Regarding the results of method A measuring the Pauli strings directly,
one can see that the measured energies are distributed
around the exact ground-state energy, and the width becomes narrower as the number of shots increases.
\par
The method B utilizing Eqs.~\eqref{eq:Hexpec}--\eqref{eq:XXsign} gives variational energy estimates, and thereby the interpretation of the results by multiple measurements is straightforward.
That is, the energy estimates by quantum circuits are given by the lowest energy among the measurements.
As the number of shots increases, results concentrate on the vicinity of the exact ground-state energies.
\subsection{Measurements of ground-state energies: noisy simulator (sim.~NISQ) case}
\par
Figure~\ref{fig:hist_NISQ} shows the sim.~NISQ results, i.e., measurements on noisy simulators.
Unlike the noise-free case, the results by method A show systematic deviations from the exact ground-state energy.
On the other hand, one can see that the method B can estimate the exact ground-state energy well even with the noisy simulator.
This is partly because the unphysical states with wrong occupation numbers are removed by the post-selection.
\par
It should be noted that the estimated energy of $ \nuc{O}{18}{} $ by method B
is distributed at slightly higher than the exact ground-state energy as shown in the panel (c) of Fig.~\ref{fig:hist_NISQ},
and the energy estimation is not necessarily improved as the number of shots are increased after $ 10^4 $.
This systematic error originates from various errors such as the gate errors, readout errors, and so on,
which are simulated based on the calibration information of the adopted real device.
For $ \nuc{He}{6}{} $, the exact ground state is the superposition
of the pair-wise configurations in $ p $ shell
and both $ p_{1/2} $ and $ p_{3/2} $ orbitals have modest contributions.
On the other hand, in the case of $ \nuc{O}{18}{} $, the exact ground state is
dominated by the pair in $ 0d_{5/2} $ and the $ 1s_{1/2} $ orbital takes the second place.
While the single-particle energies of the $ 1s_{1/2} $ and $ 0d_{5/2} $ orbitals are relatively close to each other, the $ 0d_{3/2} $ orbital with the largest single-particle energy has a large gap from those and thereby only small occupation.
In such systems having large gaps in single-particle energies,
estimating energies would be more susceptible to measurement errors in the relative amplitudes of the configurations.
Even if a small amount of extra occupation leaks into $ 0d_{3/2} $ due to measurement errors
or other reasons, one may underestimate the binding energies.
From this point of view, the result of $ \nuc{Ca}{42}{} $,
where the dominant contribution is the one occupying the $ 0f_{7/2} $ orbital
and the gaps in the single-particle energies are larger, is expected to be more sensitive to the noise.
In Table~\ref{tab:SPEs}, we summarize the single-particle energies (SPEs)
of the adopted conventional shell model interactions.
As is already mentioned, $1s_{1/2}$ and $0d_{5/2}$ orbitals in the $sd$ shell are close to each other, and $0f_{7/2}$ orbital in the $pf$ shell is the lowest in energy and having large gap from the others.
\par
In summary, for the systems of interest, the method B is superior to the method A
because it is more robust against noise.
We achieved the accuracy of at worst $ 0.5 \, \% $ with the method B on the NISQ simulator.
Besides, the method B gives the narrow distribution of the energy estimates,
demonstrating that the present method is not only accurate but also precise.
Hence, we will use the method B, utilizing computational basis sampling technique~\cite{Kohda_Qunasys_PRR22}, for real device experiments in the following.

% Table of SPEs
\begin{table}[tb]
  \begin{ruledtabular}
    \caption{Single-particle energies (SPEs) of the adopted conventional shell model interactions.}
    \label{tab:SPEs}
    \begin{tabular}{lrrr}
      interaction (nucleus) &  orbital & SPE (MeV) \\
      \hline
      CKpot ($ \nuc{He}{6}{}$)  & $0p_{1/2}$ &  2.4190 \\
                         & $0p_{3/2}$ &  1.1290 \\
     \hline
      USDB ($ \nuc{O}{18}{} $)  & $1s_{1/2}$ & -3.2079 \\
                         & $0d_{3/2}$ &  2.1117 \\
                         & $0d_{5/2}$ & -3.9257 \\
     \hline    
      GXPF1A ($ \nuc{Ca}{42}{} $) & $1p_{1/2}$ & -4.1370 \\
                         & $1p_{3/2}$ & -5.6793 \\
                         & $0f_{5/2}$ & -1.3829 \\
                         & $0d_{7/2}$ & -8.6240
    \end{tabular}
  \end{ruledtabular}
\end{table}
\subsection{Measurements of ground-state energies on real devices: Real case}
\label{sec:RealQC}
\par
We performed several measurements of the ground-state energies on real devices,
for $ \nuc{He}{6}{} $, $ \nuc{O}{18}{} $, and $ \nuc{Ca}{42}{} $.
Among the five independent experiments, the error of 
$ \nuc{He}{6}{} $ was at most $ 0.1 \, \% $,
and the typical errors for $ \nuc{O}{18}{} $ and $ \nuc{Ca}{42}{} $
were about $5 \, \% $ and $ 10 \, \% $, respectively.
Since the $\nuc{He}{6}{}$ results are as accurate as the ones on simulators,
let us focus on $\nuc{O}{18}{}$ and $\nuc{Ca}{42}{}$
and discuss the origins of the errors.
\begin{figure}[tb]
  \centering
  \includegraphics[width=1.0\linewidth]{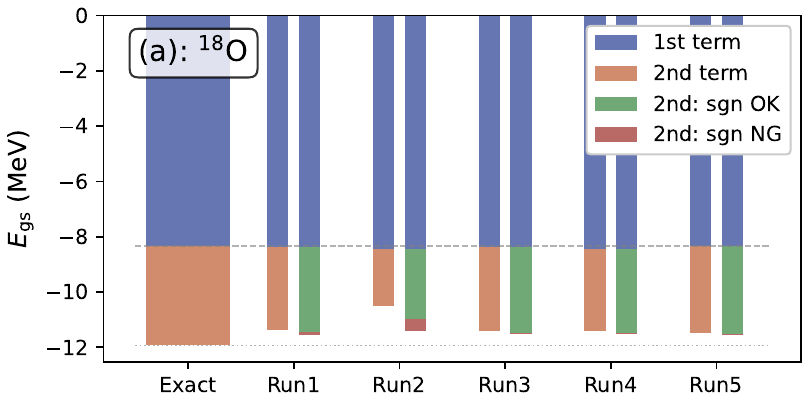}
  \includegraphics[width=1.0\linewidth]{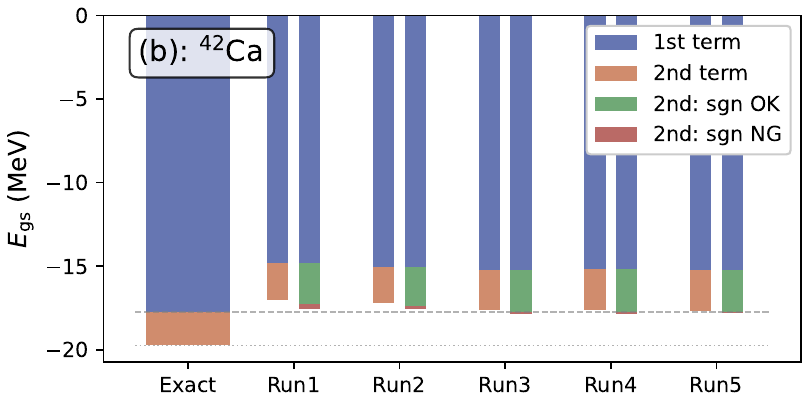}
  \caption{
    The breakdown of contributions to the energy, (a) for $ \nuc{O}{18}{} $ and (b) for $ \nuc{Ca}{42}{} $.
    The dashed and dotted lines correspond to the exact energy expectation values from
    the first (``1st'' in legend) and second (``2nd'') term of Eq.~\eqref{eq:Hpw_qubit}, respectively.
    The ``sgn OK/sgn NG'' in the legend means whether the sign of the second term was measured correctly or not,
    and the following relation holds: $ E_{\urm{2nd term}} = E_{\urm{2nd:~sgn OK}} - E_{\urm{2nd:~sgn NG}} $.
  }
  \label{fig:O18Ca42_breakdown}
\end{figure}
\par
In Fig.~\ref{fig:O18Ca42_breakdown},
the exact result and the five different results on real devices 
are plotted, and the breakdown of the contributions to the binding energy is shown by different colors.
Since the target Hamiltonian can be diagonalized
with a classical computer, one can obtain the reference value for the breakdown as shown in the ``Exact'' column.
The blue bar denote the expectation value of the first term of Eq.~\eqref{eq:Hpw_qubit}, and the orange one is for the second term.
For Run 1--5 results, the breakdown is shown in two ways.
The left ones (colored in blue and orange) are the same breakdown as the exact results.
Besides, the right bars (in green and red) show additional breakdown of the second term of Eq.~\eqref{eq:Hpw_qubit},
ones where measurements of the sign of $ XX $ and thereby $ YY $ term was correct (sgn OK in the legend) and not (sgn NG).
Here \textit{correct} means that the sign of the measured $ XX $ and $ YY $ terms was consistent with the exact and FTQC result.
Fixing these signs by hand, the measured energies were to be the lower ends of the red bars in Fig.~\ref{fig:O18Ca42_breakdown}.
The first term of Eq.~\eqref{eq:Hpw_qubit} is determined by the occupation numbers of each pair-wise configuration,
i.e., having single-particle nature,
and the second term is affected by 
both occupation numbers and the relative phase factors among the pair-wise configurations.
This breakdown illustrates some origins of the errors in the measurements of the ground-state energies.
\par
For the $ \nuc{O}{18}{} $ case, the error has multiple origins.
At first glance of Fig.~\ref{fig:O18Ca42_breakdown}, one may conclude that 
the first term is moderately estimated, and only the second term is the  source of the underestimation of the binding energy.
However, this is not the case. Looking at Fig.~\ref{fig:O18Ca42_breakdown}, we can see that
the contribution colored in red, ``2nd: sgn NG'', is small.
This means that the error comes not from the measurement of the sign of $ XX $ and $ YY $ terms
but from the amplitudes of each configuration.
It can be seen from Fig.~\ref{fig:O18_psi2} that
the occupation numbers leak from $ 0d_{5/2} $ (green) orbitals
to $ 1s_{1/2} $ (blue).
This behavior can be attributed to the fact that the occupation of the $ 0d_{3/2} $ (orange) orbital is small and rather well estimated even on the real device,
and the single-particle energies of the $ 1s_{1/2} $ and $ 0d_{5/2} $ orbitals are relatively close to each other, $ \approx 0.7 \, \mathrm{MeV} $.
In short, the origin of the error in the $ \nuc{O}{18}{} $ case is
the misestimation of the relative amplitude
between $ 1s_{1/2} $ and $ 0d_{5/2} $ orbitals.
\par
On the other hand, the error in the $ \nuc{Ca}{42}{} $ case is apparently
dominated by the first term in Eq.~\eqref{eq:Hpw_qubit},
and the error from the second term is minor.
This can be understood from, for example, Fig.~\ref{fig:Ca42_psi2},
where the ``Real'' result corresponds to ``Run1'' in the panel(b) of Fig.~\ref{fig:O18Ca42_breakdown}.
Under the adopted effective interaction, the ground state of $ \nuc{Ca}{42}{} $ is 
dominated by the configuration occupying the $ 0f_{7/2} $ orbital.
The underestimation of the first term is due to the leaking of $ f_{7/2} $
occupations to $ p_{1/2} $, $ p_{3/2} $ and $ f_{5/2} $ orbitals
having large single-particle energy gap from the $ f_{7/2} $ orbital.
Simultaneously, different experiments show underestimation of the contribution
from the first term of Eq.~\eqref{eq:Hpw_qubit}, leading to the large error in the ground-state energy.
\section{Mitigation of measurement error by circuit modification}
\label{sec:Error}
\begin{figure*}[tb]
  \centering
  \includegraphics[width=0.8\linewidth]{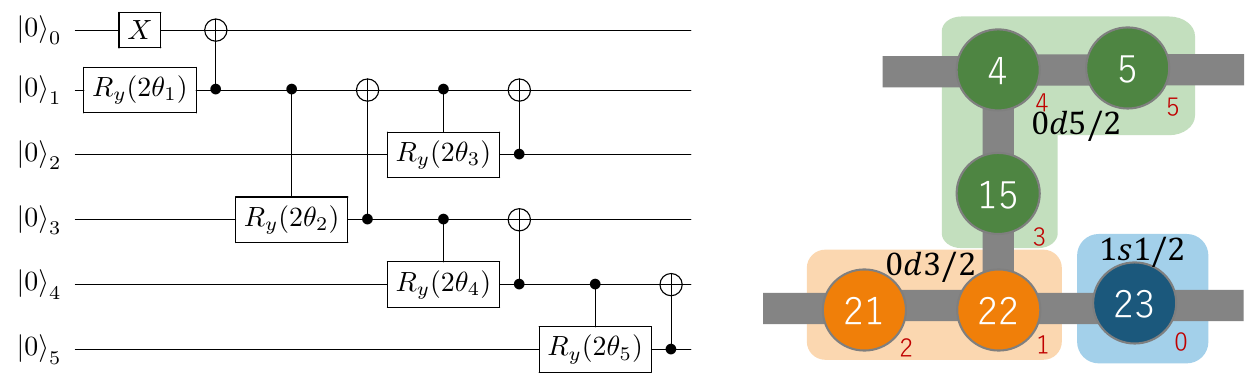}
  \caption{Example of the circuit for $ \nuc{O}{18}{} $ mitigating the error (left side),
    and a possible layout of qubits on the \textit{ibm\_brisbane} device (right side).
    The numbers such as 4, 5, 15, 21, 22, and 23 in the right side figure are the qubit indices of the \textit{ibm\_brisbane} device,
    and the small numbers associated with them are the qubit indices in the circuit.
  }
  \label{fig:O18_mitigated}
  \centering
  \includegraphics[width=0.9\linewidth]{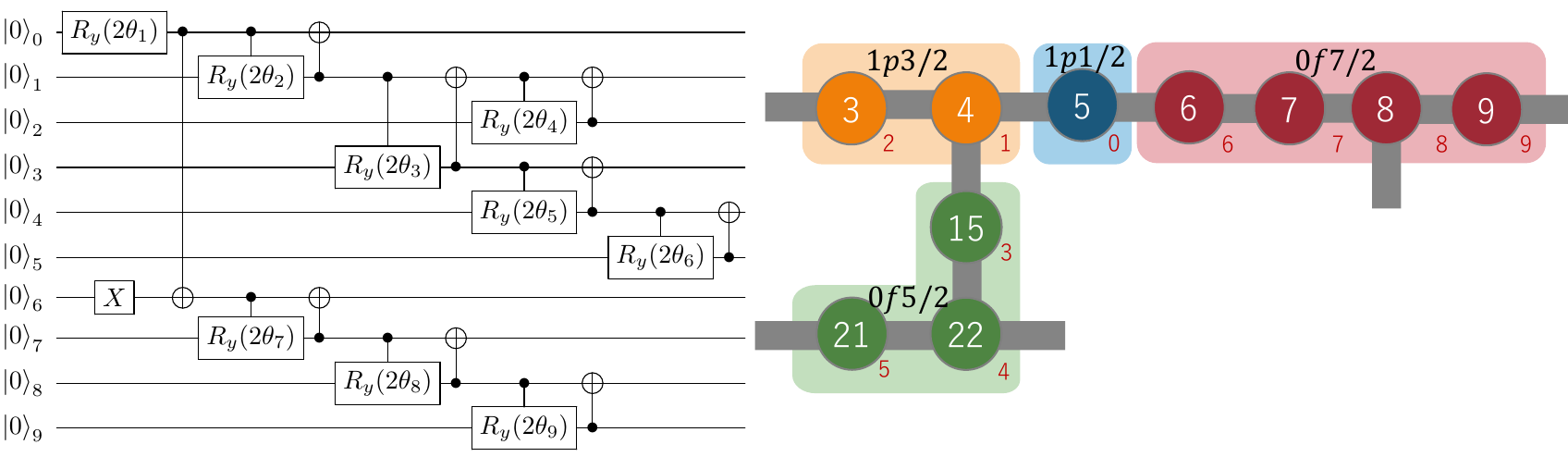}
  \caption{Counterpart of Fig.~\ref{fig:O18_mitigated} for $ \nuc{Ca}{42}{} $.
  }
  \label{fig:Ca42_mitigated}
\end{figure*}
\par
In the previous subsections, we found that the errors in the amplitudes of each configuration
is one of the major sources of the underestimation of the binding energy.
In this section, we consider a possible error mitigation strategy for the current target systems by reducing the circuit depth while keeping the equivalency of the circuit to the original one.
We will slightly modify the circuit for the wave function
and demonstrate that the accuracy of the results can be improved by this modification.
Although the method here is not an exhaustive trial of error mitigation techniques on the market~\cite{Cai_RMP23},
we will show below that it is sufficiently effective for the systems of interest.
\par
The circuit used up to this point, Fig.~\ref{fig:circuit_He6}--\ref{fig:circuit_O18}, had a structure that is to be mapped to
an unbranched chain of qubits.
Let us consider the equivalent circuits for $ \nuc{O}{18}{} $ and $ \nuc{Ca}{42}{} $ with less circuit depth.
Another criterion when modifying the circuit is that the quantum circuit
should match the layout of the adopted quantum device.
In real quantum hardware, e.g. IBMQ devices adopted in this study,
the qubits have the so-called heavy hexagon structure with the $ T $-connectivity.
If the mapping is not suited to the layout of the adopted device, one needs auxiliary qubits to implement the desired circuit.
In such cases, it is expected that the error increases due to the decoherence of the qubits involved.
\par
In Figs.~\ref{fig:O18_mitigated} and \ref{fig:Ca42_mitigated}, the modified circuits are shown in the left side.
For $ \nuc{O}{18}{} $, the circuit is modified so that the relative weights
between the different jj-coupled orbitals are calculated in earlier stage of the circuit.
Then, the weights and phases within the same $jj$-coupled orbitals are calculated.
This would prevent the leakage of the number of occupancies,
such as between $ 1s_{1/2} $ and $ 0d_{5/2} $ orbitals,
and thereby improve the accuracy of the results.
As shown in the right side of Fig.~\ref{fig:O18_mitigated},
such layout is achieved on real devices like the \textit{ibm\_brisbane} device in a straightforward manner.
This redesigned circuit turns out to improve the accuracy of the ground-state energy significantly.
The ratio of valid shots, i.e., obtaining the configurations satisfying the particle number conservation,
are increased from $ 60 \, \% $ to $ 90 \, \% $, and typical error of the energy is reduced to about $ 0.1 \, \% $.
This value is even better than the results using the original circuit with the noisy simulator, shown in Fig.~\ref{fig:hist_NISQ}.
In fact, the results of the modified circuit on the noisy simulator become comparable to or better than $ 0.1 \, \% $.
\par
For $ \nuc{Ca}{42}{} $, the circuit is modified as shown in Fig.~\ref{fig:Ca42_mitigated}.
Unlike the $ \nuc{O}{18}{} $ case, one cannot map the circuit to a chain of qubits
that determines the relative weights of the four orbitals simultaneously.
Therefore, we consider the circuit to determine the relative weights between the $ 0f_{7/2} $ orbital and the $ 1p_{1/2} $ first,
and then the relative weights within the jj-coupled orbitals
are evaluated as in the $ \nuc{O}{18}{} $ case.
By doing this, we were able to increase the percentage of the valid shots from $ 30 \, \% $ to $ 50 \, \% $ and thereby reduce the energy error to about $ 1 \, \% $.
\par
\begin{figure}[tb]
  \centering
  \includegraphics[width=1.0\linewidth]{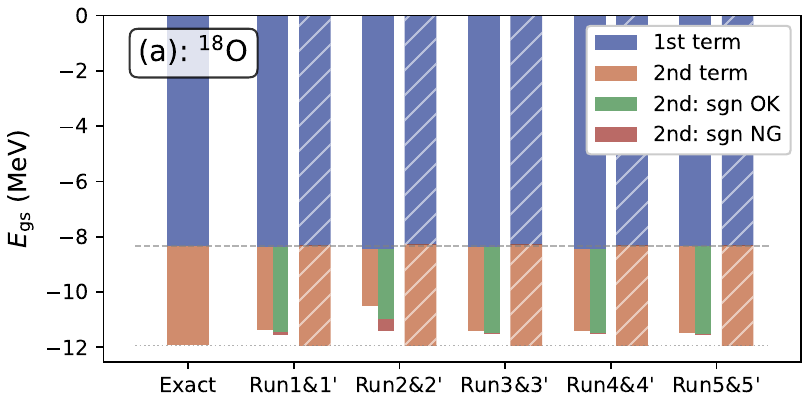}
  \includegraphics[width=1.0\linewidth]{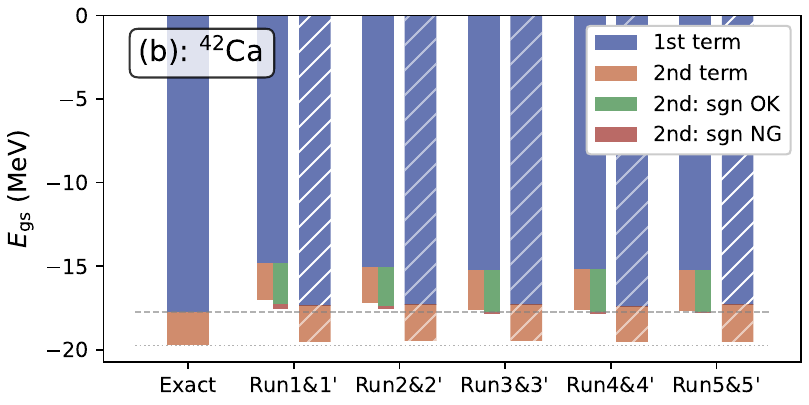}
  \caption{
    Breakdown of independent experiments for $ \nuc{O}{18}{} $ (a) and $ \nuc{Ca}{42}{} $ (b) with and without error mitigation.
    The ledends and the breakdown of the contributions are the same as Fig.~\ref{fig:O18Ca42_breakdown},
    and the hatched bars correspond to the results with the modified circuits.
  }
  \label{fig:O18_Ca42_breakdown_wwo}
\end{figure}

The impact of this modification are shown in Fig.~\ref{fig:O18_Ca42_breakdown_wwo}.
The results with modified circuits, drawn by the hatched bars, are compared with the original ones, Fig.~\ref{fig:O18Ca42_breakdown}.
One can see that the measurement errors are mitigated,
and results become much closer to the exact values.
\par
It must be noted that five different runs with and without the error mitigation,
plotted by bars side by side, are completely independent experiments.
Hence, the comparison of each pair is not meaningful.
The important point is that the error of the energy is significantly reduced by modification of the circuit layout,
and the improvement is not accidental,
but obtained consistently in all the independent experiments.
\par
The error mitigation methods we are considering now are oriented to the system under consideration,
but the results here imply that it is important, as naturally expected,
to explicitly consider symmetries and layouts of qubits
in order to obtain accurate results on NISQ devices.
\section{Summary and Outlook}
\label{sec:Summary}
\par
In this work, we have proposed an ansatz for the ground state of valence two-neutron systems
and its application to $ \nuc{He}{6}{} $, $ \nuc{O}{18}{} $, and $ \nuc{Ca}{42}{} $
on noise-free and noisy simulators and real quantum devices.
\par
For simulator results, we have shown that the proposed ansatz
with the computational basis sampling technique~\cite{Kohda_Qunasys_PRR22}
gives accurate and precise measurements of the ground-state energies of these systems.
\par
Regarding the results on real devices, we have seen some deviations
from the estimated results on noisy simulators for $ \nuc{O}{18}{} $ and $ \nuc{Ca}{42}{} $.
We analyzed the origin of this error and it turned out that the error comes
largely from the misestimation of the amplitudes (occupation numbers) of each configuration.
Then, we proposed an error mitigation technique to improve the accuracy of the results.
By considering the symmetry of the system and the layout of qubits on real quantum devices, 
we have designed the modified circuits that yield more accurate measurements
for $ \nuc{O}{18}{} $ and $ \nuc{Ca}{42}{} $ on the \textit{ibm\_brisbane} device.
For the $ \nuc{O}{18}{} $ system, we have shown that the redesigned circuit, 
which calculates the relative weights between different jj-coupled harmonic oscillator orbitals earlier in the circuit, 
leads to a significant improvement in the accuracy of the results.
The typical error of the energy was reduced to about $ 0.1 \, \% $.
Similarly, for the $ \nuc{Ca}{42}{} $ system, we have modified the circuit to firstly determine 
the relative weights of the $ 0f_{7/2} $ and the others to decrease the number of the circuit layers.
This modification increased the ratio of the valid shots and reduced the energy error to about $ 1 \, \% $.
\par
By considering these factors, we have achieved accurate measurements 
of ground-state energies for two-neutron systems.
Our results demonstrate the importance of explicitly considering symmetries 
and qubit layouts in obtaining accurate measurements,
especially on near-intermediate-scale quantum devices.
\par
The target systems discussed in this work are two-neutron systems in the valence space, which are rather simple.
The extension of this work to the systems with more valence nucleons is important, and is left for future work.
For multi-neutron systems, in particular even number systems as in~\cite{Sarma_PRC23,Bhoy_2402.15577},
the UCC ansatz can be rather simplified using the Givens rotation,
which is essentially to consider only the excitations and deexcitations
of the time-reversal pairs of the single-particle states,
and that is shown to be a good approximation for Oxygen isotopes on the $sd$ shell.
Even for that case, the number of CNOT gates is still large,
and the error for the 2-neutron system ${}^{18}$O is about 3$\%$ on the IonQ device~\cite{Sarma_PRC23}.
Our direction, exploring a system-oriented ansatz, is complementary to the works utilizing the UCC ansatz.

The ``pair-wise'' form used in this work has been used in quantum chemistry referred to as the ``antisymmetrized geminal power (AGP)''
to be used as a reference state of the coupled cluster calculation for the strongly correlated systems
to describe two-particle correlation more properly than a single Slater determinant~\cite{
  Coleman1965J.Math.Phys.6_1425,
  Uemura2015Phys.Rev.A91_062504,
  Uemura2019Phys.Rev.A99_012519,
  Khamoshi2019J.Chem.Phys.151_184103,
  Dutta2020J.Chem.TheoryComput.16_6358, 
  Khamoshi2020QuantumSci.Technol.6_014004,
  Khamoshi2021J.Chem.Phys.154_074113,
  Khamoshi2023QuantumSci.Technol.8_015006},
where a geminal is a two-body correspondence of a one-body orbital.
The operator of the geminal is quite similar to the Bardeen-Cooper-Schrieffer (BCS) one
and, actually, it is pointed out in Ref.~\cite{Khamoshi2019J.Chem.Phys.151_184103}
that the AGP wave function is strongly related to
the particle-number conserving BCS wave function~\cite{Dietrich1964Phys.Rev.135_B22}.
As an example, Ref.~\cite{Khamoshi2023QuantumSci.Technol.8_015006} discusses
the AGP wave function beyond zero-seniority configurations.
Therefore, it is natural to use this AGP ansatz for even-even nuclei as a starting point
since the valence nucleons of even-even nuclei form Cooper pairs~\cite{
  Bohr1958Phys.Rev.110_936}.
It should be noted that the ground-state wave function of the BCS state can also be written by a Pfaffian, instead of as Slater determinant~\cite{
  Bajdich2006Phys.Rev.Lett.96_130201,
  Bajdich2008Phys.Rev.B77_115112,
  Mizusaki2012Phys.Lett.B715_219};
a Pfaffian wave function and an AGP one are also based on the same philosophy.
The number of degrees of freedom involved is different between nuclear physics and quantum chemistry,
but the extension of AGP-like ansatz and the techniques developed in quantum chemistry
to nuclear many-body problems is an interesting direction to explore.

At the same time, it is interesting to test various hardwares and error mitigation techniques available in the market.
The execution time and the decoherence time of NISQ devices are trade-off, i.e.,
superconducting devices like IBM machines adopted in this work have the faster execution time
but the shorter decoherence time, while the trapped-ion devices have the slower execution time but the longer decoherence time.
To know the features of each device facilitates for further exploration of the suitable circuits,
error mitigation techniques, and the algorithms for the systems of interest.

\begin{acknowledgments}
  We acknowledge the use of IBM Quantum services for this work.
  We thank Dr. Enrico Rinaldi and Prof. Osamu Sugino for the fruitful discussion.
  This research was supported in part
  by JSPS Grant-in-Aid for Scientific Research (Grant Nos.~JP19H00869, JP20H05670, JP21K18903, JP22H05025, JP22K14030, JP22K20372, JP23H01845, JP23H04526, JP23K01845, JP23K03426, JP24K17057), 
  JST COI-NEXT (Grant No.~JPMJPF2221), 
  and MEXT Q-LEAP (Grant No.~JPMXS0118067246).
  S.Y.~acknowledges JGC-Saneyoshi Scholarship Foundation.
  T.N.~acknowledges
  the RIKEN Special Postdoctoral Researcher Program.
\end{acknowledgments}
\appendix
\section{Optimization of circuit parameters}
\label{sec:Optim}
\par
In this work, we used the fixed circuit parameters, which were obtained to reproduce the exact results.
However, for future quantum simulations addressing nuclear many-body problems,
which necessitate leveraging the full capabilities of quantum computing,
obtaining exact circuit parameters through classical computing methods will be unfeasible.
Consequently, it becomes imperative to investigate methods for optimizing the parameters in the circuit,
especially in relation to the chosen ansatz.
\par
In this section, using noise-free/noisy simulator, we show that
the proposed circuit for two-neutron systems can be optimized through measurements.
Of course, this is not full optimization of the circuit parameters using real devices,
so it is only a partial verification of feasibility of optimization,
but it provides encouraging results for future applications.
\par
We followed the optimization method proposed in Ref.~\cite{NFT_PRR20}.
The loss function, which corresponds to the expectation value of the Hamiltonian,
is written in the following form:
\begin{equation}
  \mathcal{L} \left( \bm{\theta} \right) 
  =
  \sum_k
  w_k
  \brakket{\phi} 
  {U^{\dag} \left( \bm{\theta} \right) 
    \mathcal{H}_k
    U \left( \bm{\theta} \right)}
  {\phi},
\end{equation}
where $ k $ is the index of the Hamiltonian terms
(such as $ I $, $ Z $, $ XX $, $ YY $ terms in Eq.~\eqref{eq:Hpw_qubit}),
$ w_k $ is the coefficient for $ k $-th term,
$ \ket{\phi} $ is the initial state,
and $ U \left( \bm{\theta} \right) $ is the unitary transformation corresponding to the circuit.
\par
Let us take the case of $ \nuc{O}{18}{} $ in Fig.~\ref{fig:O18_mitigated}, as an example.
For the first parameter $ \theta_1 $, which is associated with the $ R_y $ gate,
the loss function becomes the form of
\begin{equation}
  \mathcal{L} \left( \theta_1 \right)
  =
  A \cos \left( \theta_1 - B \right) + C.
\end{equation}
Then, the optimal value of $ \theta_1 $ is determined
by three measurements of $ \mathcal{L} \left( \theta_1 \right) $ at e.g., $ B $, $ B \pm \pi/2 $.
For the other parameters $ \theta_j $ ($ j > 1 $), which are associated with the controlled-$R_y$ gates,
we can write down the loss function to be minimized in the following form (see, Sec.~II D of Ref.~\cite{
  NFT_PRR20}):
\begin{equation}
  \mathcal{L} \left( \theta_j \right)
  = A \cos \theta_j +  B \sin \theta_j  +  C \cos 2\theta_j +  D \sin 2\theta_j + E.
\end{equation}
By measuring five different points, $ \theta_j $, $ \theta_j + \pi/5 $, \ldots, $ \theta_j + 8\pi/5 $,
the coefficients can be calculated via discrete Fourier transformation and the parameter, which is to be explored next, can be easily determined.
\begin{figure}
  \centering
  \includegraphics[width=1.0\linewidth]{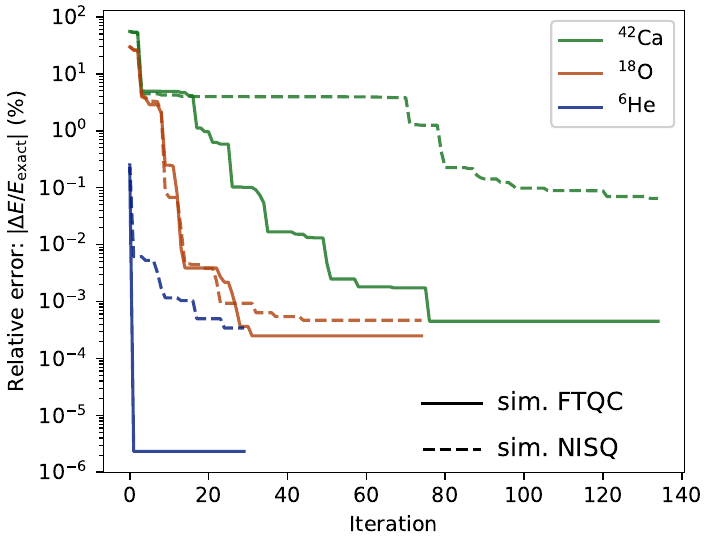}
  \caption{
    Results of optimizing the circuit parameters on noise-free and noisy simulators.
    The numbers of parameters are 2, 5, and 9 for $ \nuc{He}{6}{} $, $ \nuc{O}{18}{} $, and $ \nuc{Ca}{42}{} $, respectively.
    Each iteration corresponds to an update of a single parameter.
  }
  \label{fig:OptPara}
\end{figure}
\par
In Fig.~\ref{fig:OptPara}, we show the results of optimization of the circuit parameters
as a function of the number of iteration,
where the $ y $-axis shows the relative errors $ 100 \times \left| \Delta E / E_{\urm{exact}} \right| \, \%$.
The noise-free case (sim.~FTQC) is drawn by the solid lines,
while the dashed lines show the results of noisy simulator (sim.~NISQ).
Both results are obtained by the same circuit with 100,000 shots starting from the same initial random parameters.
Since the source of errors for sim.~FTQC is only the statistical error,
the large number of shots leads to the accurate results.
On the other hand, the sim.~NISQ case has other sources of errors,
such as the gate errors and the readout errors of the quantum device.
Note that the results for $ \nuc{O}{18}{} $ and $ \nuc{Ca}{42}{} $
are given by the circuits with the error mitigation discussed in Sec.~\ref{sec:Error}.
It should be noted that the results of the sim.~NISQ case depends not only on the circuit,
but also on the calibration status of the quantum device at the time of the experiment.
\par
Each iteration corresponds to the update of a single parameter in the circuit.
For each nucleus, 15 sweeps of optimization were performed and
all parameters are optimized only once per sweep.
The order in which the parameters are selected for updating is chosen randomly.
The sim.~NISQ results are worse than the sim.~FTQC cases by a few orders of magnitude as expected,
but one can see that the optimization of the circuit parameters successfully reduces the error and 
finally reaches acceptable errors, less than $ 0.1 \, \% $.
%
%%%%%%%%%%%%%%%%%%%%%%%%%%%%%%%%%%%%%%%%%%%%%%%%%%%%%% 
\bibliography{ref}
%%%%%%%%%%%%%%%%%%%%%%%%%%%%%%%%%%%%%%%%%%%%%%%%%%%%%% 
\end{document}